# tFold-TR: Combining Deep Learning Enhanced Hybrid Potential Energy for Template-Based Modeling Structure Refinement

Liangzhen Zheng, Haidong Lan, Tao Shen, Jiaxiang Wu, Sheng Wang, Wei Liu, Junzhou Huang*

Tencent AI Lab, Shenzhen 518057, China

*Correspondence: Junzhou Huang
Email: joehhuang@tencent.com

## Abstract

Protein structure prediction has been a grand challenge for over 50 years, owing to its broad scientific and application interests. There are two primary types of modeling algorithms, template-free modeling and template-based modeling. The latter one is suitable for easy prediction tasks and is widely adopted in computer-aided drug discoveries for drug design and screening. Although it has been several decades since its first edition, the current template-based modeling approach suffers from two critical problems: 1) there are many missing regions in the template-query sequence alignment, and 2) the accuracy of the distance pairs from different regions of the template varies, and this information is not well introduced into the modeling. To solve these two problems, we propose a structural optimization process based on template modeling, introducing two neural network models to predict the distance information of the missing regions and the accuracy of the distance pairs of different regions in the template modeling structure. The predicted distances and residue pairwise-specific deviations are incorporated into the potential energy function for structural optimization, which significantly improves the qualities of the original template modeling decoys.

## Author Summary

Protein structures are fundamental information for understanding protein functions and biological processes. While experimental protein structure determination methods are time-consuming and expensive, protein structure prediction methods are faster and high-throughput, and they are of great value to obtain near-experimental structure predictions. As one of the most commonly used and successful algorithms in the early days for structure prediction, the template-based modeling method is often used to generate homologous protein models for drug screening or functional analysis but has two significant drawbacks. We focus on solving the problems in the template-based modeling method by complementing the missing template information predicted by a deep learning model and introducing more dynamic restraints by considering the template-based modeling local quality. The refinement protocol greatly improves the template-based modeling results. This method may be further improved by incorporating a more accurate template-based modeling result quality assessment method and would benefit protein homology modeling, function analysis, and drug design.

# 1. Background

Proteins participate in almost all the essential biological activities of organisms. Structural information is vital to understand protein functions. Since the 1960s, several experimental methods, such as X-ray spectroscopy[1,2], Nucleic Magnetic Resonance (NMR)[3–5], small-angle X-ray scatter[6–8], and Cryogenic Electron Microscopy (cryo-EM)[9–12], have been developed to determine protein structures. These experimental methods are time-consuming and expensive; thus, they restrict the high throughput applications for protein structure determination in biological sciences and the pharmaceutical industry. The constraints of experimental methods and the accessibility of large-scale computational power thus promote the rise of computational methods for protein structure predictions.

Computer-aided protein structure prediction is based on a naive assumption: protein structure is uniquely determined by its amino acid sequence[13–15]. The assumption then could be further extended to another one: proteins with similar sequences could share a common folding pattern. These assumptions, therefore, urge us to consider a protein structure prediction method based on sequence identical or similar proteins whose structures have been determined. Template-based modeling (TBM)[16–23] was developed as the most pioneering protein structure prediction algorithm.

Common TBM pipelines are composed of three major steps: template searching, template-query sequence alignment, and three-dimensional (3D) modeling. Firstly, by sequence similarity searching (BLAST[24,25] or HHblits[26]) among the sequences of structure determined proteins (in Protein Data Bank), a template (which has the highest sequence similarity score to the query protein sequence) is selected. Secondly, the template sequence is aligned with the query sequence to generate the template-query alignment. The corresponding coordinates of backbone atoms (in the well-aligned region of the alignment) in the template protein structure thus are duplicated to serve as the initial coordinates for the aligned residues in the query protein for modeling. In contrast, the residues in the unaligned region in the alignment thus would be modeled randomly. Finally, the initial conformation of the query protein is then subjected to energy minimization[27–30], or Monte Carlo (MC) simulation, or molecular dynamics (MD) simulations[28,31,32], or a combination of them, guided by a particular potential energy function[33,34] ($E(\mathbb{R})$) (Equation 1), which contains the general molecular mechanics (MM)[35,36] (such as CHARMM force field[37–40] and Rosetta Ref2015 energy function[17]) and distance/contact/orientation constraints[41–47]. The constraints used for structure refinements are Cβ-Cβ (or Cα-Cα) inter-residue distance constraints as depicted by $g(d_{i,j})$ in Equation 1, $L$ indicates the length of the query sequence, and $i$ and $j$ are the indices of the residues, $\mathbb{R}$ represents the conformation of the query protein in three-dimensional (3D) space. Lastly, the collected conformations from 3D-modeling are ranked[33,34,48,49] or clustered[50] to select the best predictions.

$$E(\mathbb{R}) = \sum_{i,j}^{L} g(d_{i,j}) + P(\mathbb{R}) \quad (1)$$

The current most popular TBM pipelines, such as RosettaCM[17], Modeller[33], SWISSMODEL[40,51], I-TASSER[52–55], and RaptorX[56–58], all adopt a more or less similar protein structure prediction solution as described above.

Another type of modeling method is template-free modeling (FM), or *de novo* prediction. Increasingly accurate deep learning models have been recently applied in FM by providing accurate residue-residue distance/orientation estimations[41,42,59,60]. These distances are further utilized as the reference points for distance constraints in $g(d_{i,j})$ for minimization and refinement. In a 2019 paper[41], Yang et al. proposed that orientations could also be adopted as structure constraints in Equation 1, and their work greatly improves the overall FM accuracy.

However, in the current TBM pipelines, especially the single-template-based pipelines, the 3D modeling steps have two major problems. The first one is that for a given query sequence, the best template often does not necessarily cover the entire query sequence. In one way, because different proteins have different lengths, even proteins of the same family may differ significantly in their terminal or loop regions. There could be regions in the query sequence without an aligned counterpart in the template sequence. In another way, in many X-ray experiments, the signals of some loops cannot be obtained, and there are defects in the PDB structure of the template, leading to missing reference coordinates for the query sequence. To deal with these two types of missing information, traditional single-template-based TBM pipelines adopted very naïve approaches by assuming no inter-residue distance constraints and leaving them rather mobile in the 3D modeling step. Although loop modeling methods have been developed to complete the loops, the accuracy is still to be improved[33,61,62].

The second problem is that the current distance constraints of 3D modeling based on single-template are often treated equally[16,17]. The accuracies of the distance pairs (extracted from the template structure) used for query protein modeling are not necessarily the same. Some regions in the template sequence are highly identical to the query sequence, the distances from these regions should be more credible, and some are less similar to the query sequence thus should be less reliable. For example, an accurate template can be found in the first half of the query sequence, but the template similarity in the second half is relatively low. Intuitively, we should consider stronger inter-residue distance constraints for the first region and weaker distance constraints for the last half. Various methods are used to estimate the strength of the distance constraint of the same position of multiple templates based on the variability of the distance of the same inter-residue distances[63–66]. However, these methods assume that multiple templates can genuinely reflect the query residue-residue distance constraint distribution, which may not be valid.

To solve the above two problems, we tried to 1) predict the inter-residue distances of all amino acids by a deep learning model to fill in the missing amino acid pair information, and 2) predict the inter-residue distance deviations of the template structure and the ground truth structure by another deep learning model to obtain the residue pairwise specific distance constraint strength. With the aid of the predicted distances and deviations, we design a hybrid potential equation that can better compensate for the shortcomings of the TBM pipelines. We call the whole structure optimization process Tencent Folding for Template-based Structure Refinement (tFold-TR) (Figure 1). Based on CASP13TBM and a CAMEO 3-month dataset, we optimized the TBM modeling decoys obtained by CNFpred's template selection and RosettaCM's 3D modeling and improved the modeling average accuracy GDT-TS from 66.8 to 74.8 and from 67.1 to 72.1, respectively. The refinement process also improves the accuracy of decoys generated by other threading and 3D modeling methods.

**2. Results**

2.1 Overall performance of tFold-TR

We compared the tFold-TR refinement performance of decoys generated by different TBM pipelines. For the CASP13TBM dataset, we used three different threading methods (CNFpred, hhblit2, and BLAST) to identify the best template and used two different 3D modeling tools (Modeller and RosettaCM) to build the decoys of the query proteins based on a single-best-template. Among the 3 threading methods, CNFpred as a profile-profile threading method provides the estimated best template for all the query sequences, whereas BLAST only predicts the templates for easy targets. As for the 3D modeling methods, with the same threading method, RosettaCM builds slightly better decoys compared to Modeller (Figure 2). For example, the decoys generated by CNFpred+RosettaCM have an average GDT-TS = 66.0, whereas the average GDT-TS of decoys built by CNFpred+Modeller is 65.1. The same trend also goes for the other two threading methods on the other two datasets.

The two strategies (fixed-delta and predicted-delta) implemented in tFold-TR further refine the results from the TBM decoys generated by different threading-modeling combinations. With the CASP13TBM dataset, when the templates are selected by CNFpred, and the decoys are comparatively modeled by RosettaCM, the average accuracy of GDT-TS=66.0 is achieved. After the *fixed-delta* tFold-TR refinement, the average GDT-TS scores of the tFold-TR refined decoys improve to 72.9. Alternatively, the TBM decoys could be refined by *predicted-delta* tFold-TR, and the resulting average GDT-TS for the 74 proteins is 74.8. Thus, these results indicate that *predicted-delta* tFold-TR could increase the modeling accuracy by refining the TBM decoys. Interestingly, if we provide the ground true structures to derive the true distance deviation matrix for *delta* tFold-TR, the average GDT-TS of the CASP13TBM dataset is 79.5, which serves as the upper boundary of the modeling accuracy without providing the ground truth inter-residue distance map. Similarly, if CNFpred+Modeller generates the TBM decoys, the average GDT-TS scores of the refined decoys by three refinement strategies (*fixed-delta*, *predicted-delta*, and *native-delta*) are 7.3, 9.5, and 15.3 higher, respectively, than the initial models.

The refinement performance is also evaluated with the other two CAMEO datasets. For the CAMEO-HY dataset, *fixed-delta* and *predicted-delta* tFold-TR algorithms also increase the average GDT-TS from 65.1 to 69.2 and 72.8, respectively, when CNFpred+RosettaCM generates the initial models. To assess the performance of tFold-TR, we collected all targets in the CAMEO platform for 3 months to form a relatively larger real-world dataset (CAMEO-3M) for evaluation. For this dataset, the average GDT-TS of initial TBM decoys (CNFpred+RosettaCM) is 67.1, and it is 70.2 and 72.1 after refinement by tFold-TR *fixed-delta* and *predicted-delta* strategies, respectively.

For all the 3 datasets, if we use different threading methods for template searching and different 3D modeling tool for decoy generation, the tFold-TR pipeline could still refine the decoys and have similar accuracy improvements (Supplementary Table). To eliminate the bias from the evaluation metric (GDT-TS), TM-score and LDDT(global) scores are also calculated, and the scores of tFold-TR are also higher than the initial models (Supplementary Table). In summary, the tFold-TR pipeline could further refine the decoys generated by different TBM pipelines.

2.2 The comparison with other protein structure prediction methods

We compared the performance of tFold-TR (CNFpred+RosCM with the *predicted-delta* strategy) with other structure prediction methods. For the 3 datasets (Figure 2, D-F), tFold-TR shows the best performance (for both GDT-TS and LDDT score) compared to RaptorX[42,43], Zhang group[67,68], Robetta[69], and AlphaFold1[59,60].

On the CASP13TBM dataset, the decoys from the Zhang group have the highest average GDT-TS score (71.97), and the decoys from AlphaFold1 have the highest average global LDDT score (64.27). By applying refinement modeling using tFold-TR *predicted-delta*, the quality of the decoys generated by CNFpred+RosettaCM improves from GDT-TS 66.0, which is at the same level as the BAKER team but lower than RaptorX-DeepModeller, to a higher score 74.75, much higher than all other groups mentioned above. The refined decoys by tFold-TR also have a higher average LDDT score (66.99). For a second dataset CAMEO-HY, we collected the modeling decoys from the other three servers (RaptorX, SwissModel, and Robetta). The initial models we adopted for tFold-TR refinement have much lower average GDT-TS = 65.1 than the three servers. However, after refinement, the quality of the decoys (average GDT-TS =

73.64) has been greatly enhanced and is the most effective method among the solutions provided by the three servers and the TBM pipeline (CNFpred+RosettaCM).

Lastly, we also compared the performance of tFold-TR with other servers on the 3rd dataset CAMEO-3M. During the 3-month submission period, there were several targets where the 3 servers submitted no decoys, so only the 171 commonly submitted protein targets were selected to make a fair comparison between different methods. SwissModel and Robetta generate better decoys, though SwissModel decoys have higher GDT-TS (71.64) scores, whereas Robetta decoys have higher LDDT (70.49) scores. Our baseline TBM pipeline built the initial models for tFold-TR refinement on 171 targets, and the average GDT-TS score is increased from 67.1 to 72.2, and the average LDDT score is improved to 69.23, which is slightly higher than Robetta.

The comparisons show that the TBM pipeline (CNFpred+RosettaCM) is the least accurate method. Nevertheless, the TBM pipeline could be further improved by the tFold-TR refinement protocol to generate much better decoys. Furthermore, tFold-TR combined with a TBM pipeline is one of the most accurate protein structure prediction methods.

2.3 The relationship between initial model quality and refinement performance

From the results obtained from the CASP13TBM dataset (Figure 3A), the improvements of GDT-TS scores refined by tFold-TR (*predicted-delta* and *fixed-delta* strategy) are much more significant for medium to hard targets whose initial TBM model qualities are less than 60.0. A similar pattern is also found in the CAMEO-HY dataset (Figure 3B), which contains mainly medium to hard targets. In contrast, for easy targets whose initial TBM decoy GDT-TS scores are higher than 80.0, the tFold-TR refinement could barely improve the TBM decoy quality.

To quantify the tFold-TR performance, we adopted a metric ΔGDT-TS to measure the model quality change after refinement. When the TBM decoys of CASP13TBM targets (Figure 4A) are built by the CNFpred+RosettaCM pipeline, the refinement performance ΔGDT-TS of tFold-TR (*fixed-delta* strategy) is negatively correlated (Pearson correlation -0.575, p-value < 0.01) with the GDT-TS scores of the initial models. Moreover, the performance of *predicted-delta* strategy based tFold-TR shows a higher correlation (Pearson correlation -0.706, p-value < 0.01) with the initial models' qualities. The relationship between tFold-TR performance and initial model qualities is also examined with targets in the CAMEO-HY dataset (Figure 4B). The Pearson correlation coefficients are -0.46 and -0.736 for tFold-TR *fixed-delta* strategy and *predicted-delta* strategy, respectively.

To summarize, tFold-TR could further improve the qualities of the TBM decoys, and the quality improvement performance is more significant when the initial decoy quality is limited.

## 3. Discussion
We propose that single-template-based modeling methods have two systematic problems. The first is that in the template-query alignment, the insertion and missing regions of the query sequence would not receive coordinate information and could not be constrained to maintain short-range and long-range distances. For example, the SwissModel[40] web server would not model the template missing regions of the query sequence in the template-query alignment. In contrast, the I-TASSER server[54] uses the ab initio method to model the template missing regions by Replica Exchange Monte simulations[70]. We believe it is necessary to restrict the nearly random modeling of the template missing regions. Therefore, in *fixed-delta* tFold-TR, we implement a distance prediction model to predict inter-residue distances and make a chimeric distance map by filling up the void bins in the TBM decoy derived distance map with the information from the corresponding positions in the predicted distance map.

The second problem is that for the mutation regions of the template-query alignment, current single-template-based methods consider only unified distance constraint strength for residue pairs. It is commonly known that different amino acids have different mutation rates. Moreover, in a template-query alignment, different mutation types should thus have different information contents. For example, if we divide the template-query alignment into different local alignment regions separated by insertion or missing loops (Figure S2), the template-query similarities could be diverse in these local alignment regions. It thus is natural to believe that a more similar local alignment means more reliable local contact interactions from the template in this region. If we extend the idea into a distance map, whether a more reliable residue-residue alignment pair would provide a more accurate distance constraint remains to be further explored. Thus, we introduced a simple alignment quality score (AQS) defined by the residue mutation scoring matrix BLOSUM62[71,72] to estimate the reliability of the residue-residue distance constraint extracted from the TBM decoy. By incorporating the AQSs into the $\delta$ value of the harmonic distance constraints in the 3D modeling by RosettaCM, the refinement performance is slightly better than *fixed-delta* tFold-TR. Next, we introduced the deviation model (*tFold-RefineNet*) to estimate the quality (distance deviations from ground truth) of the distance map extracted from the TBM initial model. Then in the *predicted-delta* strategy tFold-TR, we provided a residue-residue pairwise specific $\delta$ value (distance deviations) in the harmonic distance constraints, which further increased the modeling accuracy.

Generally speaking, *fixed-delta* tFold-TR is introduced to tackle the template missing problem, while *predicted-delta* tFold-TR directly utilizes the residue-residue pairwise specific $\delta$ values to refine the TBM decoys to solve the second problem. In the following sections, we demonstrate the detailed performance gain from these two strategies.

3.1 It is necessary to provide distance constraints for template-missing regions

In the 3D modeling step of a TBM pipeline, the distance map is extracted from the template to serve as stationary distance points in a series of distance constraints. Thus, the collection of distance constraints serve as the coarse-grained energy functions to guide gradient-based minimization or Monte Carlo simulations. In both Modeller and RosettaCM, the distance constraints are defined by a harmonic form potential function with two parameters ($\delta$ and d0), where d0 is the reference stationary distance extracted from template structure and $\delta$ works as a constrain strength modulator. A larger $\delta$ value, therefore, would pose a weak distance constraint between a pair of residues' C$\alpha$ atoms (or C$\beta$ atoms). Since it is impossible to extract coordinates for the missing residues (Figure S2) in the template, whose distances for other residues thus could not be determined either, this results in deficient qualities for these residues.

It is clear that when the ratio of the missing template residues is high, indicating the target protein is a hard case for modeling, the quality of the TBM decoy tends to be low, and the ratio is indeed negatively correlated (Pearson correlation coefficient -0.588, p-value < 0.01) with initial model quality (GDT-TS score) (Figure 5A). If we feed the modeling process with deep learning predicted distances in the template missing regions, proteins with more missing regions receive considerable model quality improvement. The quantitative evidence is that the tFold-TR (*fixed-delta* strategy) performance Δ GDT-TS positively correlates (Pearson correlation coefficient 0.656, p-value < 0.01) with the ratio (Figure 5B).

The local modeling qualities (defined by local LDDT scores) of the template missing residues (or called unaligned residues) in the TBM decoys are much lower than other residues (aligned regions and mismatched residues) (Figure 5C, 5D, and Figure S2) in the dataset CASP13TBM. For the TBM (CNFpred+RosettaCM pipeline) decoys, the peak of the local

LDDT distribution is around 76.0 for the aligned residues and 35.0 for the unaligned residues. The same goes for the decoys generated by the other two TBM pipelines (HHpred+RosettaCM and BLAST+RosettaCM). The distributions of the local LDDT scores of residues in different regions indicate that the unaligned residues are poorly modeled in single-template-based modeling pipelines. Furthermore, this evidence suggests that the modeling qualities of the unaligned residues have a much larger scope for improvement.

Then tFold-TR *fixed-delta* strategy is used to refine the TBM decoys, and the distributions of the local LDDT scores of the unaligned residues shift right, indicating that, compared to the initial models, the residues in the unaligned regions in the refined decoys are better modeled. Meanwhile, the refinement does not affect the local qualities of the residues in aligned regions and mismatch regions (Figure 5C and 5D).

Taking target protein T0986s1-D1 as an example (Figure 6), we demonstrate the necessity of providing constraints for the unaligned regions. CNFpred was utilized for template searching and template-query sequence alignment, and 4dsdA was identified as the best template, and the overall sequence identity of the alignment is only 8.7%. In the template-query sequence alignment, the N-terminal region of the template is not aligned with the query sequence, and this template missing region consists of 48.3% of the whole query sequence. At the same time, the C-terminal half (residue index 40-93) contains many mismatch residues. It is not surprising that the best template is a very weak one since target T0986s1-D1 is classified as a TBM/FM target. Based on the single-template (4dsdA), the initial model (blue) generated RosettaCM protocol has a GDT-TS=32.9, and its overall folding pattern is quite divergent from the native structure (gray).

tFold-TR (*fixed-delta* strategy) further refined the initial model to a much better 3D model (GDT-TS=75.8), whose overall folding pattern is well aligned with the native structure, with the helices and sheets in the correct position and orientation. The ΔLDDT(local) scores, defined by local LDDT scores of the refined model subtracting the scores of the initial TBM model, range from 8 to 62. The residues in the N-terminal half have an average ΔLDDT(local) score around 40.0, while the average ΔLDDT(local) of the C-terminal half of the target protein is less than 30.0. The two peaks (residues 73-79 and 89-93) in the ΔLDDT(local) bar plot coincident with the two beta-sheets. The model from the *predicted-delta* tFold-TR has the highest global quality (GDT-TS=81.2). The local qualities of the N-terminal residues also have been improved, and most of the residues in the C-terminal half have higher ΔLDDT(local) scores than those in the decoy model generated by *fixed-delta* tFold-TR.

All in all, a higher ratio of template missing (unaligned) residues would result in lower quality of the TBM initial model, which could be further refined using tFold-TR by incorporating predicted distances constraints to yield a high-quality 3D model.

3.2 The relationship between alignment quality and modeling accuracy

From the traditional TBM perspective, high-quality template-query alignment is a crucial input for 3D modeling[16,17]. However, current modeling methods do not consider the information that different regions in the alignment would contribute differently to the modeling process for a single-template pipeline. It is also not clear how to introduce local alignment quality into the modeling.

To further explore the relationship between local alignment quality and local modeling quality, we firstly generate a hypothesis: if two residues are both from highly identical (high quality) alignment regions, then the distance of the two residues ($i$ and $j$) determined by the corresponding residues in the template structure would also be more reliable, meaning that the distance deviation from the ground true distance in the native structure should be close to 0). Thus the local qualities of residue $i$ and $j$ in the TBM decoy should also be relatively high (without considering the short-range and long-range interacting residues). If this hypothesis holds, then residue $i$ and $j$ should be "trusted" more than other residues by implementing a stronger constraint (smaller $\delta$ value) on the distance between these two residues in the energy function.

To prove that the hypothesis makes sense in some cases, a local alignment quality score (AQS) of residue $i$ ($q_i$) is defined by averaging the alignment direct scoring ($s_{i-k}$ to $s_{i+k}$) based on the BLOSUM62 matrix. For example, if residue $i$ is a Threonine (Thr) and the corresponding residue $i$ in the template is also a Thr, then $s_i$ equals 6. If residue $i$ in query sequence is a gap open, then -11 is assigned to $s_i$. Alternatively, if residue $i$ in query sequence is a gap extent, $s_i$=-1. The local alignment quality score of residue $i$ is the average of the direct scoring of the neighboring residues (from residue $i-k$ to $i+k$).

The statistics support the claim that residues with low AQSs have relatively lower modeling qualities in CASP13TBM and CAMEO-HY datasets. For the targets in these two datasets, the CNFpred+RosettaCM pipeline was used for template searching, template-query sequence alignment, and 3D modeling to obtain the best initial models, where both the local AQS scores of the alignment and the local structure quality scores (local LDDT) of the decoys were examined for the mismatched residues.

Clearly, the residues with "good" alignment quality (higher ASQ scores) are naturally modelled by TBM pipeline with "good" structure quality (higher local LDDT scores), and the Pearson Correlation Coefficients are 0.446 (p-value < 0.01) and 0.558 (p-value < 0.01) for the CASP13TBM and CAMEO-HY datasets. There is a strong correlation between the local alignment quality and local structure quality. Thus it would be reasonable to trust the residues with high local alignment quality during the 3D modeling, and the hypothesis proposed would be partially tested.

To use the local alignment quality information in the 3D modeling, especially the $\delta$ values in the harmonic constraints, we input the algorithmic average of the AQS of residue $i$ and $j$ into to impact the constraint strengths. However, the trial is not very successful since the average GDT-TS scores of the decoys generated by refinement modeling only slightly increased. The problem is that we need a direct estimation of $\delta_{ij}$, but we used the average of $q_i$ and $q_j$; the information may not be direct enough.

Nevertheless, it is still an open question remaining to be further explored. Instead, in this work, we are equipped with *tFold-RefineNet* to probe the problem directly, estimating $\delta_{ij}$ by predicting the distance deviation ($dev_{ij}$) between the distance of residue $i$ and $j$ in the decoy and the distance in the native structure to transform the $dev_{ij}$ into $\delta_{ij}$. The introduced deviation matrix provides a pairwise-specific $\delta$ value prediction and thus directly shapes the coarse-grained potential energy function in the 3D modeling.

Furthermore, tFold-TR improves both local modeling qualities (Figure 7C, 7D, and Figure S2) and global modeling accuracies (Figure 2A-2C). For the residues with low local AQS in the alignment, their local structure qualities (in the initial model) are also low, but after tFold-TR refinement, the local structure quality improvements are more significant. Although the correlation between AQS and ΔLDDT(local) is not very strong for the CASP13TBM dataset (-0.184) and CAMEO-HY dataset (-0.415).

Taking T0958-D1 as an example (Figure 8), the initial 3D model of CNFpred+RosettaCM (GDT-TS=38.6) has poor accuracy, the proportion of missing regions in alignment is 11.7%, and the sequence similarity is 13.0%. The accuracy of the 3D structure after optimization by *fixed-delta* strategy reaches 50.0, and the overall modeling accuracy of the second missing region (31-41) has improved significantly. However, the position of the two β-sheets at the C-terminus of the model is not maintained. In the native structure, there are three β-sheets: residues 20-24, residues 60-65, and residues71-75. The first and the third should form strand pairing, and this structural pattern is not recovered in both the initial model and the optimized model by *fixed-delta* tFold-TR. Moreover, in the model optimized by the *predicted-delta* strategy, the overall accuracy reached 74.7. The RMSD against the native structure decreases from 12.53 Å to 2.40 Å in the initial model and refined model, while the relative positions of these three β-sheets are closer to the experimental structure, as can be seen in Figure 8, the ΔLDDT(local) scores of the three β-sheet regions in are much higher in *predicted-delta* tFold-TR generated structure.

Another example, T0993s2-D1 (Figure S4), is also from CASP13TBM. The similarity of the alignment of T0993s2-D1 is relatively high, and there are no missing regions. The overall accuracy of the 3D structure obtained by CNFpred+RosettaCM for this target is GDT-TS=71.9, and the overall folding is consistent with the experimental structure. After optimizations by the two strategies of *fixed-delta* and *predicted-delta* by tFold-TR, the accuracy (GDT-TS) of the 3D structure is improved to 76.3 and 81.6, respectively.

In summary, we employ pairwise-specific restraints to model the mismatch regions, and this strategy (*predicted-delta*) works better than the *fixed-delta* strategy, partially solving the second drawback of TBM modeling mentioned earlier.

3.3 Continuous improvement of distance deviation prediction is meaningful

We believe that there is room for continued improvement in the structural optimization of the *predicted-delta* strategy. First, with a known native structure, the average accuracy of the initial models from the CNFpred+RosettaCM pipeline is GDT-TS=66.0 (CASP13TBM dataset), which can be optimized to 79.8 by the *native-delta* strategy (Supplementary Table). This observation suggests that our current prediction model for deviations has room for improvement, and by improving the prediction accuracy of deviation, we may be able to continuously improve *predicted-delta* tFold-TR performance. Interestingly, we observed a negative correlation between the deviation prediction error RMSE (in Å unit) and the optimization performance of the *predicted-delta* strategy with the CASP13TBM dataset (Figure 9). When the RMSE is less than 3 Å, the accuracy of the refined decoy models with multiple targets is greatly improved (ΔGDT-TS > 20). Therefore, it is a potential research direction to continuously improve the prediction accuracy of distance deviation for more accurate template-based decoy refinement.

4. Methods

4.1 Datasets

We collected 3 datasets (from CASP13 and CAMEO platforms) to build TBM decoys and refine them with our tFold-TR pipeline.

The first dataset (CASP13TBM) contains all 73 domain-level protein sequences for modeling in CASP13 TBM and TBM/FM categories. The second dataset (CAMEO-3M) contains 74 medium-to-hard proteins collected from the CAMEO 3D modeling platform from September 2019 to May 2020. The selection criteria are that the Cα LDDT score of the decoy predicted by server 999 (single-best-template server) should be less than 40, and the sequence length of a protein should be between 50-300 to avoid too long or too short sequence targets. A third dataset (CAMEO-HY) was also collected from CAMEO, and it is composed of all 182 query sequences used by CAMEO online server predictions during a half-year period (27th Jun. 2020 to 19th Sept. 2020).

For dataset 1, we collected prediction results from Zhang[46] (group 322) and RaptorX (group 324) from https://predictioncenter.org/casp13. As for datasets 2 and 3, we downloaded the predictions from three servers, RaptorX[42,43], Robetta[69], and SwissModel[40], respectively. All the ground-truth structures of the sequences in the 3 datasets were downloaded from the RCSB Protein Data Bank (PDB)[73]. The target IDs and sequences can be found in the Support Information.

Table 1. The query sequence datasets used to evaluate refinement performance.

| SN | Dataset Name | Description | # Sequences | Date Range |
|---|---|---|---|---|
| 1 | CASP13TBM | Casp13 TBM and FM/TBM | 73 | 2018.3 to 2018.5 |
| 2 | CAMEO-HY | Cameo half year hard targets | 74 | 2019.12.7 to 2020.5.30 |
| 3 | CAMEO-3M | All targets in cameo latest 3 months | 182 | 2020.6.27 to 2020.9.19 |

4.2 Databases used for MSA searching

For the 3 datasets in Table 1, we used different databases for template searching and MSA searching to feed the distances and deviation prediction deep learning models, as listed in Table 2. The databases are set up here to ensure that there are no data leaks. For example, for both CASP13TBM and CAMEO-HY datasets, we used the databases before the CASP13 competition, while for the CAMEO-3M dataset, we used the November 2019 and earlier versions of the database for all our MSA searching procedures.

Table 2. The databases involved in this paper and the corresponding stamps of them

| Database Name | CASP13TBM | CAMEO-HY | CAMEO-3M |
|---|---|---|---|
| Template database | 2018.03 | 2018.03 | 2019.07 |
| Uniclust30 | 2017.10 | 2017.10 | 2018.08 |
| UniRef90 | 2018.03 | 2018.03 | 2019.07 |
| NR | 2018.04 | 2018.04 | 2019.11 |
| MetaClust50 | 2018.01 | 2018.01 | 2018.06 |

4.3 Template searching and TBM decoy generation

tFold-TR takes the predicted inter-residue distance map, the inter-residue deviation map, and a TBM generated decoy as inputs to compose a hybrid potential energy function to build better decoys. For each sequence to be queried, we performed a template search using different threading methods (CNFpred[74], hhblits v2.0[26], and BLAST[24,25]), and we finally selected the top-ranked one as the single template for modeling. For the CASP13 and CAMEO-HY datasets, their template search was performed against the same template sequence library (PDB non-redundant database before March 2018), while the CAMEO-3M dataset used a somewhat newer template database. The default parameters were used in CNFpred's template search, i.e., a p-value cutoff of 0.005, and *mainscore* was used as the final ranked score. The search parameters for hhblits is e-value 0.001 with 3 iterations, while the other parameters are set as the default values. The template search parameters for BLAST are e-value < 0.01, coverage over 40%.

Ultimately, for the three datasets, the number of targets for which the three different threading methods succeeded in finding the templates varied (Table below). For example, for some hard targets, BLAST and hhblits failed to find useful templates given the searching parameters. The template-sequence alignments were generated by respective threading methods. Visualization of Alignment was done with the online server AlignmentViewer [75] (https://alignmentviewer.org/).

Table 3. The number of targets for which different threading methods can find templates.

| Dataset Name | CNFpred | HHblits | BLAST |
|---|---|---|---|
| CASP13TBM | 73 | 60 | 34 |
| CAMEO-HY | 74 | 59 | 49 |
| CAMEO-3M | 182 | 158 | 135 |

At the first stage, decoy production is performed according to single-best-template and template-query alignment, both using the standard RosettaCM protocol and Modeller simulation protocol. For each input sequence, each 3D modeling method (RosettaCM or Modeller) produces 300 decoys, and the best decoy is selected as input for the subsequent optimization round. In this case, the best decoy generated by RosettaCM is the one with the lowest Rosetta Energy (considering the constraint part). The best decoy from Modeller is chosen by Spicker[50] to select the center structure of the cluster with the largest population.

4.4 Distance and deviation prediction

The distance prediction is based on a distance prediction model (*tFold-DistNet*) developed by our group, the input features of the model are extracted from Multiple Sequence Alignment (MSA), i.e., Position-Specific Scoring Matrix (PSSM)[76] and Markov Random Fields (MRF)[77,78], through recurrent convolutional layers, it can finally output the inter-residue distances, the inter-residue orientation, as well as backbone dihedral angle and secondary structure prediction. The MSAs were generated by different MSA search tools (hhblits2[26], HMMER[79], and BLAST[24,25]) at different e-values for several databases (Uniclust30, Uniref90, NR, and MetaClust50 as listed in Table 2). Different distance predictions can be obtained by different input MSAs, and the best distance prediction is obtained by clustering and other filtering methods. In addition, we used a 2$^{nd}$ deep learning model (*tFold-RefineNet*) for template decoy inter-residue distance deviation (from the native structure) prediction. The deviation is defined as the absolute difference of the Cβ atom distance between a pair of residues from the TBM modeled decoy and the native structure. For a query sequence with length $L$, a TBM pipeline would generate a pool of decoys and select the best decoy (initial model) as the input for the above deviation model, which yields an $L \times L$ deviation matrix. The values in the deviation matrix range from -10Å to 10Å. We use the $L \times L$ predicted decoy deviation matrix to determine the strength of the distance constraints for all residue-residue pairs. The two deep learning models will be explained in another paper by our group.

4.5 3D modeling in tFold-TR

For common TBM and FM pipelines, in the 3D modeling phase, they are trying to restrict the distances of all residue pairs as the primary constraints to construct the potential energy equation to find a global minimum. Here, we consider incorporating the distance prediction from *tFold-DistNet* and the TBM decoy to make a hybrid potential energy function to refine the TBM decoy to generate more accurate 3D models. The following text explains the core idea of how we modify the potential energy functions for TBM decoy refinements.

Based on a threading method for template screening and query-template alignment and a 3D modeling approach, for any pair of residues $(i, j)$, a constraint $g$ (in Equation 1) is imposed to maintain its distance in the desired range around the distance extracted from template structure. Generally, for 3D modeling in TBM pipelines (such as RosettaCM), a simple harmonic function (see Equation 2) is used for distance constraints,

$$g(i,j) = \frac{(d_x - d_0)}{\delta} \quad (2)$$

where $d_x$ represents the distance between residue $i$ and $j$ during the modeling, while $d_0$ is the optimal distance for residue $i$ and $j$ calculated based on template structure. $\delta$ is a constant for constraint strength adjustment, a larger $\delta$ would enable a weak constraint, and a smaller $\delta$ would impose strong restraints for the distance between residue $i$ and $j$. The essence of the hybrid potential energy function introduced by tFold-TR is the modification of $\delta$ and the definition of $d_0$: 1) for residue pairs without reliable distance from the template structure; the distance model predicts the distances to complete the missing puzzle; 2) for each residue pair, instead of a constant $\delta$, a residue-residue pairwise specific $\delta$ is introduced to account for the variable reliabilities of the distances ($d_0$) extracted from the template structure. Here, in the tFold-TR pipeline, the best decoy from the TBM pipeline is used as the *initial structure model* for refinement, which serves as the new "template structure" to extract respective distances between residues.

More specifically, a new equation for constraints is then designed as the following:

$$g(d_{i,j}) = \begin{cases} g_{FM}, & \text{if } align(i,j) = 0 \\ g_{TBM}, & \text{if } align(i,j) = 1 \end{cases} \quad (3)$$

For residues $i$ and $j$ in a TBM decoy, if they both could find aligned counterpart residues in the query-template alignment ($align(i,j)=1$, in the aligned regions of a query-template alignment, Figure S2), their distance reference point $d_0$ is the distance of the two aligned counterpart residues (Cβ-Cβ) in the template (*initial model*). Furthermore, the distance constraint energy function takes precisely the same formulation as Equation 2, except we used a variable $\delta_{ij}$ to replace the original constant $\delta$. The "HARMONIC" type of constraints[17] defined by Rosetta is used to restrict the distance between residue $i$ and $j$.

However, if either one of residue $i$ or $j$ could not find a counterpart aligned residue in the query-template alignment (in the unaligned regions, Figure S2), we set $align(i,j)$ as 0, and the distance reference $d_0$ is then taken from the distance map predicted from the distance model. The constraint energy function is then defined as a spline type formed by the logarithm of predicted probabilities of the discrete distances as used by trRosetta. For a spline constraint of the Cβ-Cβ, a probability threshold ($p_{cut}=0.2$) is set to determine whether to use this constraint. In addition, all optimization processes used orientations[41] between residues based on deep learning predictions.

To fully evaluate the effects of different $\delta$ settings, we adopted 3 strategies for TBM decoy refinement in tFold-TR. For *fixed-delta* tFold-TR, we set $\delta$ to 5, which is carefully selected to have the best decoy refinement performance with the CASP13TBM dataset. For *predicted-delta* tFold-TR, the constraint energy function adopts a unique $\delta_{i,j}$ for a specific residue pair $i$ and $j$. Given the TBM decoy, the deviation model predicts a deviation matrix, where each bin contains a predicted inter-residue distance deviation $dev_{i,j}$ between the decoy and the native structure, the $\delta_{i,j}$ is transformed from the absolute value of $dev_{i,j}$ (Equation 5). The third strategy is set to verify where the theoretical upper limit of the modeling accuracy of our optimization algorithm is if the deep learning method can predict the true deviation with 100% accuracy. In this strategy, $dev_{i,j}$ is the absolute value of the difference between the Cβ atom distance of the residue $i$ and $j$ of the native structure and the decoy.

Table 4. The three strategies deployed with tFold-TR

| Strategy Name | $\delta$ setting | How $\delta$ is determined |
| --- | --- | --- |
| *fixed-delta* | $\delta = 5.0$ | A hand-craft parameter |
| *predicted-delta* | $\delta_{i,j}$ | Deviation model, Equation 4 and 5 |
| *native-delta* | $\delta_{i,j}$ | Deviations between TBM decoy and native structure, Equation 4 and 5 |

$$g_{TBM}(d_{i,j}) = (\frac{d_{i,j}-d_0}{\delta_{i,j}})^2 \quad (4)$$

$$\delta_{i,j} = \begin{cases} 1.0, & if\ dev_{i,j} \leq 1.0\ \text{Å} \\ abs(dev_{i,j}), & if\ 1.0\ \text{Å} < abs(dev_{i,j}) \leq 10.0\ \text{Å} \\ 10.0, & if\ dev_{i,j} > 10.0\ \text{Å} \end{cases} \quad (5)$$

tFold-TR's structural optimization starts with the best decoys obtained from RosettaCM or Modeller. The optimization protocol is based on PyRosetta: inputting a TBM decoy as the initial model, starting with multiple rounds of energy minimization in centroid mode, followed by a full-atom fast-relax, and then clash removal. For one TBM initial decoy, 50 decoys are generated from tFold-TR protocols, and the decoy with the lowest energy (Rosetta energy + constraint energy) is then selected as the final refinement result.

4.6 Evaluation metrics

In the TBM step, and the later structural refinement using tFold-TR, only one decoy (with the best energy) was selected for the performance analysis.

We use different metrics for decoy quality evaluation. Taking a decoy as a whole, we evaluate the similarity between the decoy and the native structure by TMscore[80], LDDT (global)[81], and GDT-TS[82]. We used DeepScore[83] to calculate GDT-TS and TMscore, and our in-house LDDT_Assess tool to perform global and local LDDT calculations. In this paper, unless otherwise stated, LDDT refers to the LDDT (global) and is an overall quality score of the decoy, while local LDDT is a local quality assessment at the residue level. The ranges of both TMscore, LDDT, GDT-TS, and local LDDT range from 0 to 100. A higher score indicates a better global or local quality of the decoy.

**Figure Captions**

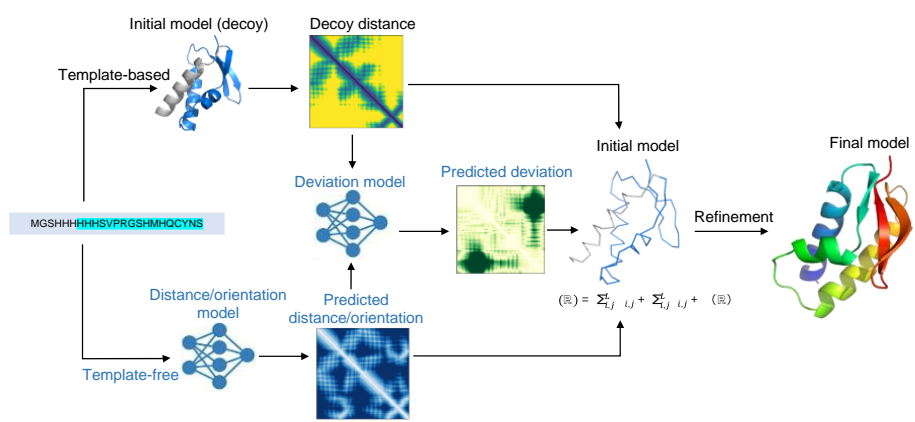

Figure 1. the tFold-TR pipeline. The query sequence is initially modeled by a TBM method to obtain the best template-based decoy (initial decoy). Next, through the distance model (*tFold-DistNet*), the distance map for the query sequence is predicted. By providing the decoy distance map and the predicted distance map, another deviation model (*tFold-RefineNet*) predicts the distance deviation matrix for the initial decoy. Combining the decoy distance map, predicted distance map, and the deviation map, a hybrid potential energy function is derived to guide the structure refinement of the initial decoy structure.

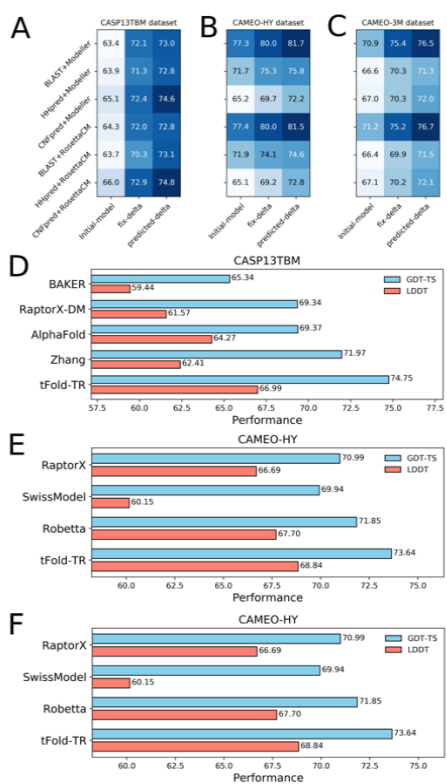

Figure 2. The performance of the hybrid potential-based approach tFold-TR for structural optimization. A-C, each bin gets a value indicating the average GDT-TS, and the darker the bin, the higher the average GDT-TS. D-E, the performance of different folding algorithms (servers) on the three datasets, where blue indicates GDT-TS and red indicates LDDT. Note that RaptorX-DM in panel D refers to RaptorX-DeepModeller, and in E, RaptorX, SwissModel,

and Robetta submitted a common set of 72 (out of 74) targets, where the performance of all methods is therefore averaged over the 72 targets. On the CAMEO-3M dataset in panel E, RaptorX skipped 88 targets, while SwissModel and Robetta have fewer missing decoys, so the performance of RaptorX is represented by the average score of the submitted 94 targets, while for SwissModel, Robetta, and tFold-TR, the performance is evaluated with the shared 171 targets.

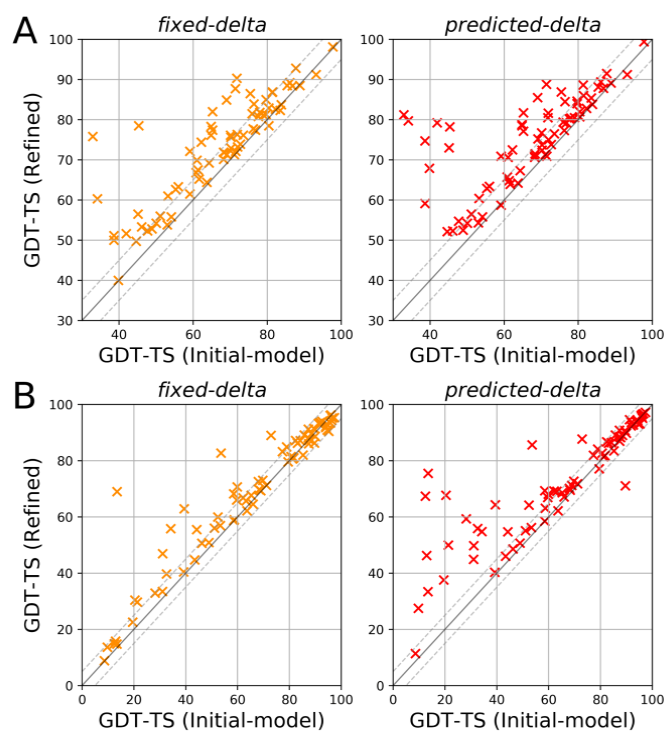

Figure 3. The performance of tFold-TR refinement of initial models for the CASP13TBM dataset (A) and the CAMEO-3M dataset (B). RosettaCM built the initial models (TBM decoys). The gray dashed lines indicate +/- 5 GDT-TS range.

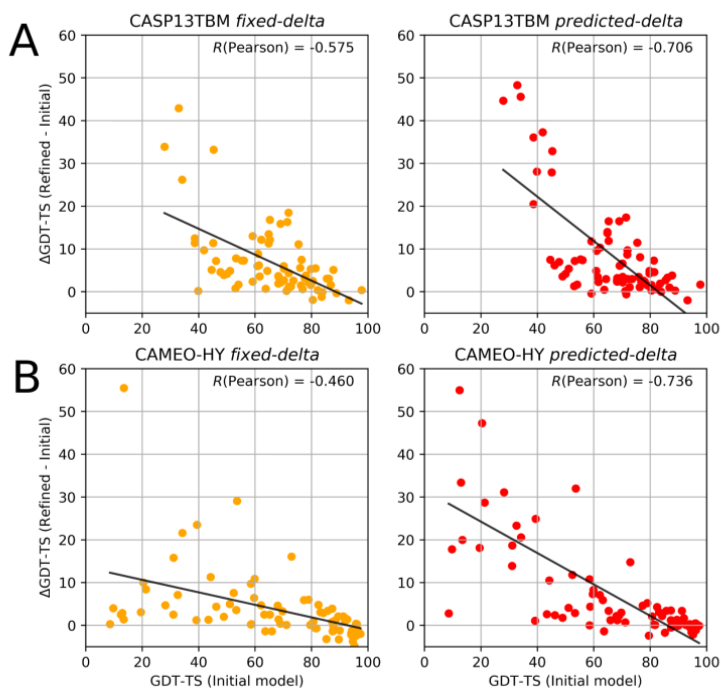

Figure 4. The effect of structural optimization is negatively correlated with the model accuracy of the initial TBM decoys for the CASP13TBM dataset (A) and the CAMEO-HY dataset (B).

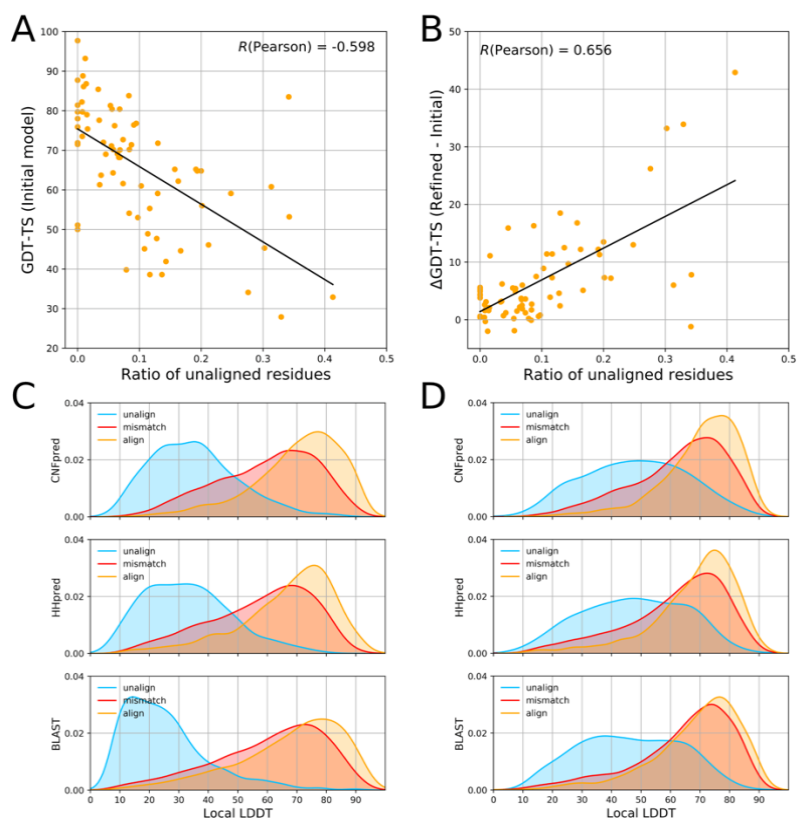

Figure 5. The proportion of unaligned regions in template-query alignment in the CASP13TBM dataset is related to the accuracy of the TBM modeling and the structural optimization by tFold-TR. (A) the relationship between the ratio of unaligned in GDT-TS and alignment of the initial model obtained by CNFpred+RosettaCM. (B) the relationship between the ratio of unaligned residues in GDT-TS and alignment after *fixed-delta* tFold-TR optimization. (C) the distribution of local LDDT in different regions of the alignment of the initial model obtained by CNFpred+RosettaCM. (D) local LDDT distribution of different regions of the alignment after *fixed-delta* tFold-TR refinement.

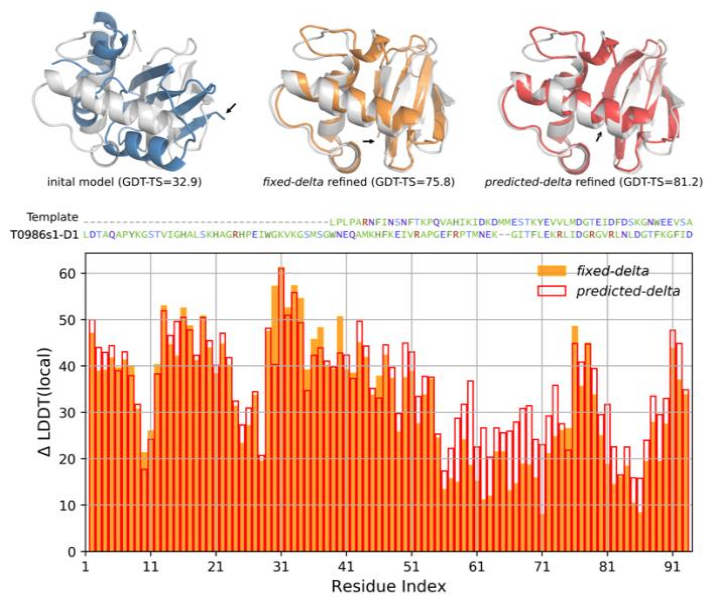

Figure 6. The refinement performance of tFold-TR for target T0986s1-D1. The native structure of T0986s1-D1 is shown as gray color in the top panels. The black arrows in the top panels indicate the C-terminal region of the domain.

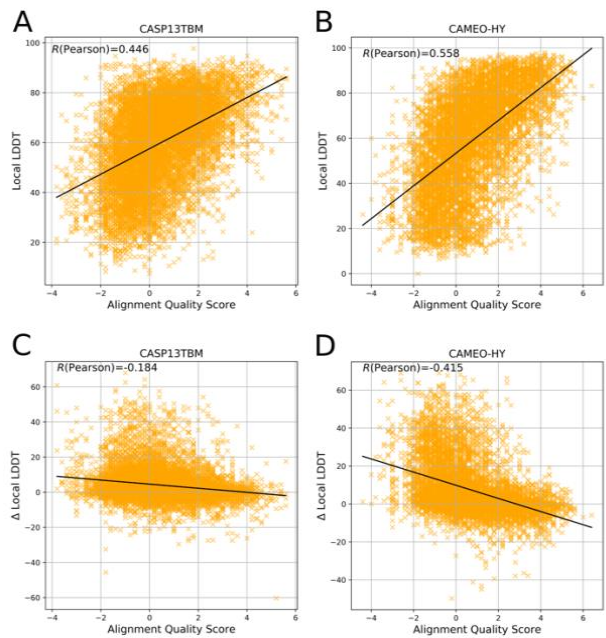

Figure 7. The relationship between the AQS of amino acids and initial model local LDDT and the accuracy improvement after structure optimization. (A) and (C), based on the CASP13TBM dataset, initial model from CNFpred+RosettaCM. (B) and (D,) based on the CAMEO-HF dataset, initial model from CNFpred+RosettaCM.

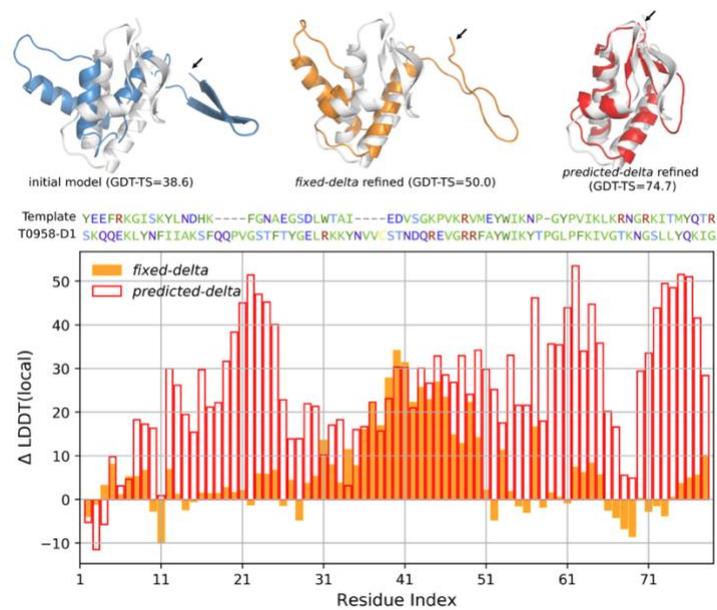

Figure 8. The refinement performance of tFold-TR for target T0958-D1. The native structure of T0985-D1 is shown as gray color in the top panels. The black arrows in the top panels indicate the C-terminal region of the domain.

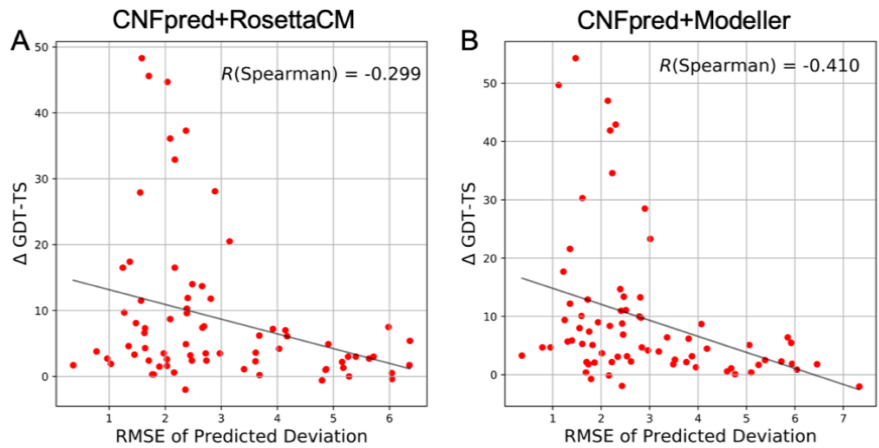

Figure 9. Relationship between distance deviation prediction error with refinement (*predicted-delta* strategy) performance. The evaluations are calculated on the CASP13TBM dataset for initial models. ΔGDT-TS is defined as GDT-TS (*predicted-delta* refined) - GDT-TS (initial model). (A), the refinement is conducted with CNFpred+RosettaCM initial models. (B), the refinement is conducted with CNFpred+Modeller initial models.

**Support Figure Captions**

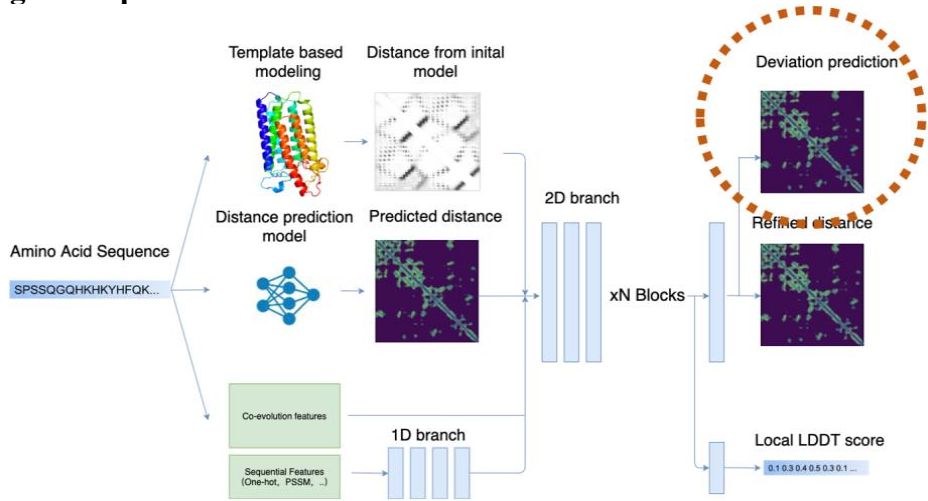

Figure S1. Deviation prediction deep learning model (*tFold-RefineNet*) pipeline.

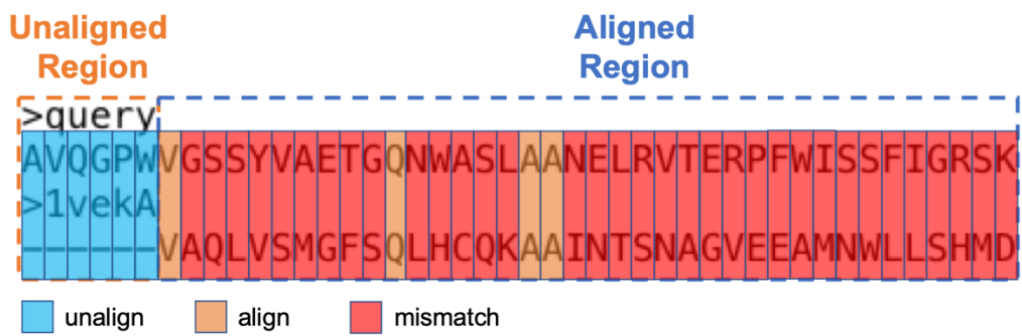

Figure S2. The different types of query-template alignment residues.

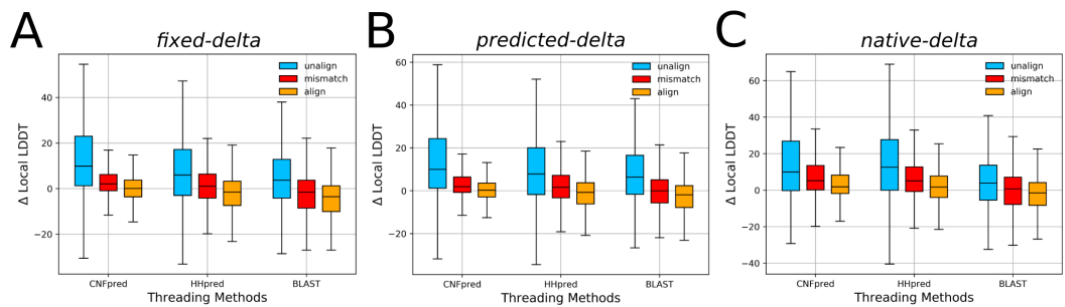

Figure S3. The accuracy increase of decoys generated by RosettaCM TBM pipelines in 3 alignment regions brought by tFold-TR with three strategies on the CASP13TBM dataset.

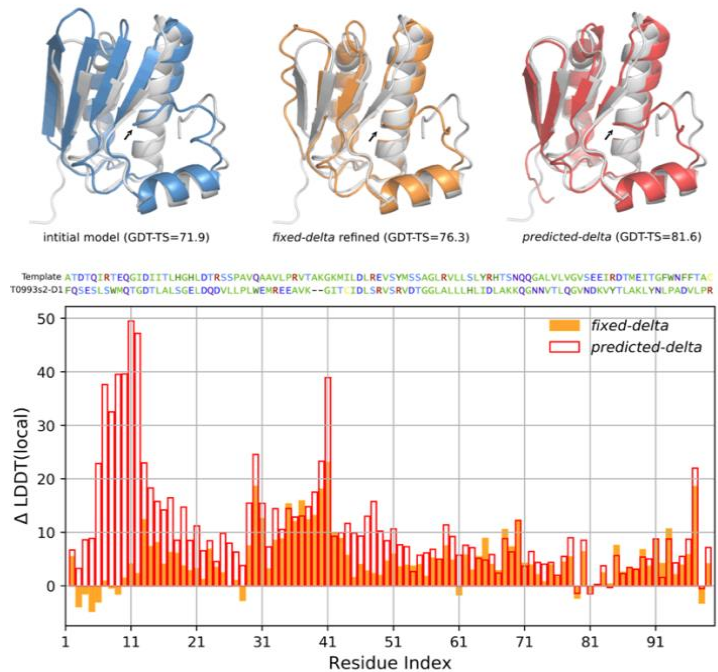

Figure S4. Structure comparison and local LDDT enhancement for T0993s2-D1 structure optimization.